\documentstyle[11pt]{article}
\newcommand{\D}{{\rm d}}

\begin{document}
\large

 {\LARGE{\bf
 \centerline { Smooth sandwich gravitational waves}}}
 \vspace{14mm}
 \centerline { J. Podolsk\' y}
 \vspace{5mm}

 {\it
 \centerline {  Department of Theoretical Physics,
           Faculty of Mathematics and Physics,}

 \centerline {   Charles University, V Hole\v sovi\v ck\'ach 2,
     180 00 Prague 8, Czech Republic}}
 \vspace{5mm}

 \centerline{{\small
    Electronic address:  podolsky@mbox.troja.mff.cuni.cz}}
 \vspace{12mm}

\begin{abstract}
Gravitational waves which are  smooth and contain two
asymptotically flat regions are constructed from
the homogeneous {\it pp}-waves vacuum solution.
Motion of free test particles is calculated explicitly and the
limit to an impulsive wave is also considered.

\vspace{3mm}
\noindent
PACS number(s): 04.30.-w, 04.20.Jb, 98.80.Hw
\end{abstract}
\vspace{14mm}

The widely known class of {\it pp}-waves \cite{KSMH} is characterized by
the existence of a quadruple Debever-Penrose null vector field which
is covariantly constant. In vacuum  the metric can be written as
\begin{equation}
\D s^2=2\,\D\zeta \D\bar\zeta-2\,\D u\D v-(f+\bar f)\,\D u^2\ , \label{E1}
\end{equation}
where  $f(u,\zeta)$ is an arbitrary function of $u$,
analytic in $\zeta$. The simplest case for which the metric
describes  gravitational waves arises when $f$ is of the form
\begin{equation}
f(u,\zeta)=d(u)\zeta^2\ , \label{E2}
\end{equation}
where $d(u)$ is an {\it arbitrary} function of $u$;
such solutions are called homogeneous {\it pp} waves
(or ``plane'' gravitational waves).
Performing the transformation \cite{Penrose}
\begin{equation}
\zeta=\frac{1}{\sqrt{2}}\,(Px+iQy)      \ ,\qquad
 v   =\frac{1}{2}\,(t+z+PP'x^2+QQ'y^2)  \ ,\qquad
 u   =t-z                               \ ,\label{E3}
\end{equation}
where real functions $P(u)\equiv P(t-z)$, $Q(u)\equiv Q(t-z)$
are solutions of differential equations
\begin{equation}
P''+d(u)\,P=0\ ,\qquad Q''-d(u)\,Q=0\ , \label{E4}
\end{equation}
(here prime denotes the derivative with respect to $u$)
the metric can be written in the form
\begin{equation}
\D s^2 = - \D t^2 + P^2 \D x^2 + Q^2 \D y^2 + \D z^2\ , \label{E5}
\end{equation}
which is more suitable for physical interpretation.
Considering free test particles
standing at fixed $x$, $y$ and $z$, their {\it relative} motion in the
$x$-direction is given by the function $P(u)$ while it is given
by $Q(u)$ in the $y$-direction. The motions are unaffected in the
$z$-direction which demonstrate transversality of the  waves.

Assuming ``profile''  functions $d(u)$ non-vanishing  on some
finite interval of $u$ only, sandwich gravitational waves were constructed
in \cite{BPR}-\cite{PV} and elsewhere. Here we
consider the  function $d(u)$ of the form
\begin{equation}
d(u)=\frac{a}{\cosh^2(bu)}   \ ,
\label{E6}
\end{equation}
where $a$ and $b$  are arbitrary real positive constants.
Since all derivatives of $d(u)$ are continuous the corresponding
gravitational waves are {\it smooth}, contrary to ``standard'' sandwich
waves explicitly presented in literature. On the other hand, the
space-times  (\ref{E1}) given by  Eqs. (\ref{E2}) and  (\ref{E6}) do not contain
flat regions in front of the wave and behind it. They
are curved everywhere, becoming flat only {\it asymptotically} as
$u\to\pm\infty$. Therefore, we should call them
``smooth asymptotic sandwich waves''.

In order to find the form (\ref{E5}) of the metric exhibiting
naturally the particle motions we solve the equations (\ref{E4}).
Introducing a substitution
\begin{equation}
\xi=-\tanh(bu)   \ ,
\label{E7}
\end{equation}
the equations take the form
\begin{equation}
(1-\xi^2)\frac{\D^2R}{\D\xi^2}-2\xi\frac{\D R}{\D\xi}\pm\frac{a}{b^2}R =0 \ ,
\label{E8}
\end{equation}
where the upper sign is applied for $R=P$ and the lower sign for $R=Q$.
Clearly, a general solution can be written as a linear
combination of the Legendre functions of the first kind $P_\alpha(\xi)$
and the second kind $Q_\alpha(\xi)$ where
$\alpha=(\sqrt{1 \pm 4a/b^2}-1)/2$. It is natural to impose the
conditions  $R(u\to-\infty)\to 1$ and
$R'(u\to-\infty)\to 0$ so that the metric (\ref{E5}) is
written in explicit Minkowski form in the asymptotic region where
$u\to-\infty$. These conditions are satisfied by particular solutions of
the form
\begin{equation}
P(u)=P_\mu(\xi(u))    \ ,\qquad
Q(u)=P_\nu(\xi(u))\ ,    \label{E9}
\end{equation}
where
\begin{equation}
\mu=(\sqrt{1 + 4a/b^2}-1)/2 \ ,\qquad
\nu=(\sqrt{1 - 4a/b^2}-1)/2 \ .\label{E10}
\end{equation}
Typical behavior of the functions $P, Q$ is shown in Fig. 1. It can
be observed that in both asymptotic regions $u\to\pm\infty$
the particles move uniformly. This can be shown
analyticly using the definition
$P_\alpha(\xi)\equiv F(-\alpha,\alpha+1;1;(1-\xi)/2)$ where $F$
is the hypergeometric function. Clearly, $P\to1, Q\to1$ as
$u\to-\infty$. For $u\to+\infty$ (corresponding to $\xi\to-1$)
we use the identity 15.3.10. in \cite{Abram} according to which
\begin{eqnarray}
F\Big(-\alpha,\alpha+1;1;\frac{1-\xi}{2}\Big) &=&-\frac{\sin \pi\alpha}{\pi}
   \sum_{n=0}^\infty \frac{(a)_n (b)_n}{(n!)^2} \label{E11}\\
&&\hskip-30mm\times\Big[2\psi(n+1)-\psi(n-\alpha)-\psi(\alpha+n+1)-
     \ln\Big(\frac{1+\xi}{2}\Big)\Big]\Big(\frac{1+\xi}{2}\Big)^n
    \ , \nonumber
\end{eqnarray}
where $\psi$ is the digamma function. Using the identity
$\psi(-\alpha)=\psi(1-\alpha)+1/\alpha$ and the limit
 $\ln[(1+\xi)/2]\to-2bu$ for  $u\to+\infty$ we
conclude that
\begin{equation}
P(u\to+\infty)=C_1+C_2u    \ ,\qquad
Q(u\to+\infty)=D_1+D_2u    \ ,    \label{E12}
\end{equation}
where
\begin{eqnarray}
&&C_1=\frac{\sin \pi\mu}{\pi}
 \Big[\frac{1}{\mu} +\psi(1+\mu)+\psi(1-\mu)-2\psi(1)\Big]    \ ,\qquad
C_2=-\frac{2b}{\pi}\sin \pi\mu\ , \nonumber\\
&&D_1=\frac{\sin \pi\nu}{\pi}
 \Big[\frac{1}{\nu} +\psi(1+\nu)+\psi(1-\nu)-2\psi(1)\Big]    \ ,\qquad
D_2=-\frac{2b}{\pi}\sin \pi\nu\ .
    \label{E13}
\end{eqnarray}
For particular values of $\mu$ corresponding to $\mu=m=0,1,2,\cdots$,
the constant $C_2$ vanishes so that $P(u\to+\infty)=C_1$. These
cases coincide with solutions $P(u)$ given by Legendre
{\it polynomials} $P_m(\xi)$ and we easily get
$C_1=P_m(\xi=-1)=(-1)^m$. Explicit particular solutions of this
type arise when $a/b^2=m(m+1)$ so that
$P(u)=-\tanh(\sqrt{a/2}\,u)$ for $m=1$,
$P(u)=[3\tanh^2(\sqrt{a/6}\,u)-1]/2$ for $m=2$, etc.
Note also that there are no analogous solutions for $Q(u)$ since
(\ref{E10}) admits $\nu\in\langle -1,0\rangle$ only. Therefore,
the only non-negative integer (for which the constant $D_2$ vanishes) is
$\nu=0$ representing a trivial case $d(u)=0$.
 For all other values of $\nu$ the constant $D_2$ is positive so
that $Q(u)\to+\infty$ as $u\to+\infty$.

We can also use the above results for construction of impulsive gravitational
waves. The sequence of smooth functions (\ref{E6}) approach the $\delta$
function (in a distributional sense) as
$a\to 0$ if the second parameter is $b=2a$ (so that the
normalization condition $\int_{-\infty}^{+\infty} d(u)\,\D u=1$
holds for arbitrary $a$). Considering this limit, indicated also
in Fig.~1,
$\alpha=(\sqrt{1 \pm 1/a}-1)/2\to 0$ and using Eqs.~(\ref{E13}) we get
\begin{equation}
P(u)=1-u\,\Theta(u)\ ,\qquad
Q(u)=1+u\,\Theta(u)\ ,\label{E14}
\end{equation}
where $\Theta$ is the Heaviside step function ($\Theta=0$ for
$u<0$, $\Theta=1$ for $u>0$).
Therefore, particle motion in the impulsive gravitational wave is the
same as in the limit of ``standard'' sandwich waves \cite{Penrose}.
\vspace{10mm}

\centerline{\bf  Acknowledgments}
\vspace{3mm}

\noindent
I acknowledge the support of grants No. GACR-202/96/0206 and
No. GAUK-230/1996 from the Czech Republic and Charles University.

\vspace{20mm}

{\LARGE Figure Caption}

\vspace{5mm}
\noindent
Fig. 1. Typical behavior of the functions $P(u)$ and $Q(u)$
determining relative motion of free test particles (initially at
rest) in $x$ and $y$-directions, respectively, caused by smooth
sandwich gravitational waves for different values of
$a=\frac{1}{48},\frac{2}{48},\cdots, \frac{12}{48}$, with $b=2a$. The
values $a=\frac{1}{8},\frac{1}{24},\frac{1}{48}$
correspond to $\mu=m=1, 2, 3$, respectively, for which $P\to(-1)^m$
as $u\to+\infty$.

\end{document}